\documentclass[aps,prb,twocolumn,showpacs]{revtex4}
\usepackage{graphicx}
\usepackage{amsmath}

\begin{document}
\title{Persistent current and Wigner crystallization in a one
  dimensional quantum ring} 

\date{\today} 

\author{Marc Siegmund}

\email{Siegmund@physik.uni-erlangen.de}

\author{Markus Hofmann}

\author{Oleg Pankratov}

\affiliation{Lehrstuhl f\"ur Theoretische Festk\"orperphysik,
  Universit\"at Erlangen-N\"urnberg, Staudtstrasse 7 B2, D-91058
  Erlangen, Germany}

\begin{abstract}
  We use Density Functional Theory to study interacting spinless electrons
  on a one-dimensional quantum ring in the density range where the
  system undergoes Wigner crystallization. The Wigner transition leads
  to a drastic ``collective'' electron localization due to the Wigner
  crystal pinning, provided a weak impurity potential is applied. To
  reveal this localization we examine a persistent current in a ring
  penetrated by a magnetic flux. Using the DFT-OEP method we
  calculated the current as a function of the interaction parameter
  $r_{\rm S}$. We find that in the limit of vanishing impurity
  potential the persistent current stays constant up to a critical
  value of $r_{\rm S}^{\rm c}\approx 2.05$ but shows a drastic
  exponential decay for larger $r_{\rm S}$ which reflects a formation
  of a pinned Wigner crystal. Above $r_{\rm S}^{\rm c}$ the amplitude
  of the electron density oscillations exactly follows the $(r_{\rm
    S}-r_{\rm S}^{\rm c})^{^{1}/_{2}}$ behaviour, confirming a
  second-order phase transition as expected in the mean-field-type OEP
  approximation.
\end{abstract}


\bibliographystyle{apsrev}

\maketitle

\section{Introduction}

In the last years, the fabrication of quasi-one-dimensional quantum
rings became possible \cite{IF03,MCB93}. In such systems only few
transverse states are occupied and by increasing the curvature of the
confining potential the system can be made effectively
one-dimensional. The number of electrons on the ring can be controlled
by the gate electrode. The experimental studies of the rings with only
one or two electrons were reported by Lorke {\it et al.} \cite{LLG00}
The possibility to vary the number of particles from very few to
several hundreds enables experimentalists to tune the
electron-electron interaction in a wide range. One of the most
striking consequences of the interaction is the formation of a Wigner
crystal \cite{W34}, a many-body state with electrons localized at
discrete lattice sites.  Yet it is well known that in an infinite
one-dimensional system the fluctuations destroy the long-range
order\cite{Landau5}. This raised doubts about the existence of a
one-dimensional Wigner crystal which has become a long-debated
subject. Only in the nineties Glazman {\it et al.} \cite{GRS92} have
shown that the arbitrarily weak pinning potential stabilizes the
one-dimensional Wigner crystal. It was proven that the pinning
potential suppresses the long-wavelength fluctuation modes which are
responsible for destroying the long-range order. Due to the pinning
potential the Wigner state is always localized in contrast to the
electron liquid state \cite{GRS92}. Thus in the presence of a weak
impurity the Wigner transition should manifest itself as electron
localization.

The critical $r_{\rm S}^{\rm c}$ for a 1D system estimated in the work
of Glazman {\it et al.} was of the order of unity. For the
two-dimensional electron gas Tanatar and Ceperley found a critical
value $r_{\rm S}^{\rm c}=37\pm 5$, using a Monte-Carlo technique
\cite{TC89}. The reason for this large value is a very small
shear modulus of the two-dimensional Wigner crystal \cite{GRS92}. In
three dimensions a Wigner crystal is expected at $r_{\rm S}>65\pm 10$
(Ref.  \onlinecite{OHB99}).

Electron localization seems to be a convenient signature to observe a
formation of the pinned one-dimensional Wigner crystal. However, in
numerical simulations it is not quite evident how to quantify the
localization of a correlated many body state. Several indirect
criteria such as the inverse participation number \cite{KK93} or the
curvature of the ground state energy \cite{Koh64} have been suggested
to distinguish between a localized and a delocalized state. However,
to the best of our knowledge, the electrons' ability to carry electric
current -- which is the most direct indication of the delocalized vs.
localized behaviour -- has not yet been explored. In this work we
calculate the persistent current in a one-dimensional quantum ring
penetrated by a magnetic flux. We apply a weak impurity potential
which pins a Wigner state but practically does not influence the
electron liquid state.

In the density range where a Wigner crystal already exists as a ground
state the persistent current has been studied analytically by Krive
{\it et al.}\cite{KSS95} for smooth potentials allowing semiclassical
treatment. In a perfect ring the Wigner crystal rotates as a whole
producing exactly the same current as non-interacting electrons.
In the presence of a weak impurity potential the persistent current
was found to be suppressed exponentially with the increasing impurity
strength or the Wigner crystal stiffness \cite{KSS95}.

We use Density Functional Theory (DFT) to calculate self-consistently
the persistent current in a one-dimensional system with ten electrons.
In the limit of vanishing (on a scale of the inter-electron Coulomb
repulsion) repulsive potential we find that the current is independent
of $r_{\rm S}$ for $r_{\rm S}<2.05$. At larger $r_{\rm S}$ the
persistent current decreases exponentially with increasing $r_{\rm
  S}$, indicating a localization of the electrons. At the transition
point the system undergoes a second-order phase transition which can
be seen by considering the amplitude of the density oscillations
$\delta$ as an order parameter. We find that in a crystalline phase
$\delta$ exactly follows a square root behaviour
$\delta\sim\left(r_{\rm S}-r_{\rm S}^{\rm c}\right)^{^{1/_{2}}}$. The
stronger pinning potentials smear the phase transition such that no
distinct transition point can be observed.

The article is organized as follows. In section II we introduce the
model of a one-dimensional quantum ring with the Gaussian impurity
potential. We briefly discuss the OEP approximation \cite{SH53,TS76}
which is used for the exchange potential. We also introduce an
Electron Localization Function \cite{BE90} which is helpful for a
real-space visualization of the electron localization. In section III
we describe the computational method for solving the self-consistent
Kohn-Sham equations. In section IV we present our results for the
persistent current as a function of $r_{\rm S}$. We consider impurity
potentials of various amplitude and width and show how these
parameters influence the current. The conclusions are given in section
V.

\section{The model}

We study a system of $N=10$ interacting spinless electrons in a one
dimensional ring of circumference $L=2\pi R$. The ring geometry is
accounted for via periodic boundary conditions and $x=\varphi R$
denotes the coordinate along the ring. A persistent current is induced
by a vector potential $\vec{A}=(A_{r},A_{\varphi})$ with a tangential
component
\begin{equation}
  A_{\varphi}=\frac{\Phi}{L} \label{vector-pot}
\end{equation}
that provides a magnetic flux $\Phi$ through the ring. The vector
potential is chosen such that the electrons move in a field-free
space.

Additionally, we introduce a repulsive Gaussian potential centered at
$x_{0}$
\begin{equation}
  V_{\rm imp}(x)=V_{0}\exp\left(-\frac{(x-x_{0})^{2}}{\sigma^{2}}\right), \hspace{3em} (V_{0}>0) \label{imp-pot}
\end{equation}
which should pin the Wigner crystal phase.

We calculate the ground state current density for a given value of the
magnetic flux and for a given strength and width of the impurity
potential using Density Functional Theory. The self consistent
Kohn-Sham \cite{KS65} equations for this system are given by
\begin{multline}
  \left[\frac{1}{2m_{0}^{\ast}}\left(-i\hbar\partial_{x}-eA_{\varphi}\right)^{2}+V_{\rm imp}(x)+V_{\rm int}(x)\right]\varphi_{i}(x)\\
  =\epsilon_{i}\varphi_{i}(x) \label{KS-eq}
\end{multline}
where index $i$ labels the Kohn-Sham orbitals $\varphi_{i}$ and the
eigenvalues $\epsilon_{i}$. The electron-electron interaction is
described by an effective one-particle scalar potential $V_{\rm
  int}=V_{\rm H}+V_{\rm OEP}^{\rm x}$. Here, $V_{\rm H}$ is the
Hartree potential and $V_{\rm OEP}^{\rm x}$ is the exchange
contribution. The latter is calculated in the KLI version
\cite{KLI92a,KLI92b} of the OEP method \cite{SH53,TS76}.

The central assumption of the OEP method is that the
exchange-correlation energy functional can be written explicitely in
terms of the Kohn-Sham orbitals. A common choice is the ``exact
exchange'' functional
\begin{equation}
  E_{\rm x}^{\rm EXX}=-\frac{1}{2}\frac{e^{2}}{4\pi\varepsilon\varepsilon_{0}}\sum_{i,j}^{N}\iint dx\,dx^{\prime}\,\frac{\varphi^{\ast}_{i}(x)\varphi_{j}(x)\varphi^{\ast}_{j}(x^{\prime})\varphi_{i}(x^{\prime})}{|x-x^{\prime}|}
\end{equation}
which has the form of the Fock energy but the wavefunctions
$\varphi_{i}$ are the Kohn-Sham orbitals rather than the Hartree-Fock
orbitals. Minimization of the full energy functional with respect to
the density leads to an integral equation for the exchange-correlation
potential. In this work we use the exact-exchange functional and apply
the KLI approximation which allows to transform the OEP integral
equation into a considerably simpler algebraic equation. Still, it
retains important features of the exact xc potential such as the
derivative discontinuities and correct asymptotic behaviour \cite{Gra00}.

Since DFT in the Kohn-Sham formulation is essentially a mean-field
theory, fluctuations are not accounted for in our calculations. It is
well known that fluctuations are particularly important in one
dimension \cite{Landau5}. But since even an infinite one-dimensional
Wigner crystal is stabilized by an arbitrarily weak pinning potential
\cite{GRS92} we expect that the fluctuations are effectively suppressed
not only due to the pinning potential, but also due to the finite size
of the ring.

Whether the ground state of a many electron system is an electron
gas-like one or a Wigner crystal state depends on the ratio of the
kinetic energy and the Coulomb energy. In one dimension this ratio is
simply proportional to the electron density $n$, whereas in two and three
dimensions it is proportional to $\sqrt{n}$ and $n^{^{1}/_{3}}$,
respectively. Hence for high densities the kinetic energy dominates
and the ground state is electron gas-like whereas for low
densities the Coulomb repulsion favours the crystalline state.

Experimentally it is most straightforward to vary the electron density
to switch between weakly and strongly interacting regimes. Yet the
variation of the electron number should alter the persistent current
even in a non-interacting system which conceals the interaction
effects. As we use a ``persistent current criterion'' to identify the
Wigner transition we prefer to exclude the aforementioned trivial
single particle contribution and to retain only the influence of
many-body effects. It can be done using an alternative (though
somewhat artificial) way of controlling the ratio of kinetic and
Coulomb energy. Namely, let us consider the effective electron mass
$m^{\ast}$ as a free parameter.  In one dimension, the energy ratio
$r_{\rm S}$ is proportional to $m^{\ast}$:
\begin{equation}
  r_{\rm S}=\frac{1}{2N}\frac{L}{a_{\rm B}}\frac{m^{\ast}}{m_{0}^{\ast}}\,{,} \label{rS}
\end{equation}
where $a_{\rm B}$ and $m_{0}^{\ast}$ are the Bohr radius and the
``true'' effective electron mass in the host material. The persistent
current density
\begin{multline}
  j(x)=-\frac{i\hbar}{2m_{0}^{\ast}}\sum_{i=1}^{N}\left[\varphi_{i}^{\ast}(x)\partial_{x}\varphi_{i}(x)-\varphi_{i}(x)\partial_{x}\varphi_{i}^{\ast}(x)\right]\\
  -\frac{\hbar}{m_{0}^{\ast}}\frac{2\pi}{L}\frac{\Phi}{\Phi_{0}}n(x)
  \label{eq:current}
\end{multline}
should be calculated with the fixed ``true'' effective electron mass
$m_{0}^{\ast}$. Here, $\Phi_{0}=\frac{h}{e}$ is the flux quantum and
\begin{equation}
  n(x)=\sum_{i=1}^{N}\varphi_{i}^{\ast}(x)\varphi_{i}(x)
\end{equation}
is the density.

Also, the ratio of the kinetic energy to the impurity potential must
be kept constant when changing $r_{\rm S}$ via changing $m^{\ast}$.
Otherwise the current density of a system of non-interacting electrons
would depend on $r_{\rm S}$. The impurity potential strength
$V_{0}$ must be renormalized as
\begin{equation}
  V_{0}\rightarrow V_{0}^{\ast}=V_{0}\frac{m_{0}^{\ast}}{m^{\ast}}\,{.} \label{potRenorm}
\end{equation}

The potential renormalization (\ref{potRenorm}) guarantees that the
artificial variation of the electron mass results in a dependence of
the persistent current on $r_{\rm S}$ solely due to the
electron-electron interaction.

Equation (\ref{eq:current}) expresses the current density via the
Kohn-Sham orbitals within the framework of the ordinary density-based
DFT. It is not, however, strictly justified since the common DFT
Kohn-Sham equations by construction yield the exact ground state
density but not the current density. Strictly speaking, one has to
employ the current density functional theory \cite{Vig88} (CDFT) which
expresses the ground-state energy functional as a functional of the
density and the paramagnetic current density. The Kohn-Sham orbitals
in CDFT thus give the exact current density of the interacting system.
However, the CDFT corrections are, as a matter of fact, usually very
small. For example in a recent paper \cite{Sharma07} is shown that the
orbital magnetic moments in magnetic (Fe, Co and Ni) and non-magnetic
(Si and Ge) solids calculated with CDFT only slightly differ from
those calculated with DFT.  Hence we expect that Eq.
(\ref{eq:current}) evaluates the current reasonably well, keeping in
mind, that for our purposes not the current value itself, but its
critical $r_{\rm S}$-dependence close to the Wigner transition is of
interest.

In addition to the current density we also use the Electron
Localization Function (ELF) \cite{BE90} to visualize the
electrons' localization. The idea behind the definition of the ELF is
that the more localized electron produces a stronger repulsion of the 
other like-spin electrons due to the Pauli exclusion principle.
According to this picture the ELF measures the probability to find a
second electron (with the parallel spin) anywhere close to a reference
electron. It is defined such that its value of one half means a
homogeneous electron-gas like state whereas a value of one
refers to a perfectly localized electron at this point in space.

In its original definition \cite{BE90} the ELF was formulated for the
real wavefunctions only. Recently it was generalized to the
time-dependent case \cite{Burnus05} where complex wavefunctions have
to be employed. This form of the ELF is also suitable 
for the current-carrying static system we consider. It is given by
\begin{equation}
  \eta(x)=\frac{1}{1+\chi^{2}(x)}
\end{equation}
with
\begin{equation}
  \chi(x)=\frac{\tau(x)-\frac{1}{4}\frac{(n^{\prime}(x))^{2}}{n(x)}-\frac{(j_{\rm p}(x))^{2}}{n(x)}}{\tau^{\rm hom}(x)}\,{.}
\end{equation}
In this expression
$\tau(x)=\frac{\hbar^{2}}{m_{0}^{\ast}}\sum_{i}|\partial_{x}\varphi_{i}(x)|^{2}$
is the kinetic energy density of the Kohn-Sham system and $\tau^{\rm
  hom}(x)=\frac{\hbar^{2}\pi^{2}}{6m_{0}^{\ast}}n^{3}(x)$ is the
respective quantity in a one-dimensional homogeneous electron gas with
density $n(x)$.

\section{Computational method}

For numerical solution of the Kohn-Sham equations (\ref{KS-eq}) we
use a real space method. We expand the wave functions $\varphi_{i}(x)$
using a spline basis \cite{HBP01}
\begin{equation}
  \varphi_{i}(x)=\sum_{\nu}a_{i}^{(\nu)}b_{\nu}(x)
\end{equation}
with the complex coefficients $a_{i}^{(\nu)}$ and the real basis functions
\begin{equation}
  b_{\nu}(x)=\left\{
    \begin{array}{r@{\quad:\quad}l}
      \frac{1}{4}\left(2+\frac{x-x_{\nu}}{h}\right)^{3} &  -2<\frac{x-x_{\nu}}{h}\leq -1\\[0.5ex] 
      1-\frac{3}{2}\left(\frac{x-x_{\nu}}{h}\right)^{2}-\frac{3}{4}\left(\frac{x-x_{\nu}}{h}\right)^{3} &  -1<\frac{x-x_{\nu}}{h}\leq 0 \\[0.5ex]
      1-\frac{3}{2}\left(\frac{x-x_{\nu}}{h}\right)^{2}+\frac{3}{4}\left(\frac{x-x_{\nu}}{h}\right)^{3} &  0<\frac{x-x_{\nu}}{h}\leq 1 \\[0.5ex] 
      \frac{1}{4}\left(2-\frac{x-x_{\nu}}{h}\right)^{3} &  1<\frac{x-x_{\nu}}{h}\leq 2\\[0.5ex] 
      0 & \textrm{else.}
    \end{array} 
  \right. \label{eq:spline}
\end{equation}
The spline nodes are $x_{\nu}$ and $h$ is the distance between the two adjacent
nodes. The basis functions (\ref{eq:spline}) are not orthogonal which means that the
overlap matrix
\begin{equation}
  S_{\mu,\nu}=\int dx\, b_{\mu}(x)b_{\nu}(x)
\end{equation}
is not diagonal. With this representation of the wave functions, the
Schr\"odinger equation reads
\begin{equation}
  \sum_{\nu}H_{\mu,\nu}a_{i}^{(\nu)}=\epsilon_{i}\sum_{\nu}S_{\mu,\nu}a_{i}^{(\nu)}
\end{equation}
with the Hamiltonian matrix
\begin{equation}
  H_{\mu,\nu}=\int dx\, b_{\mu}(x)\hat{H}b_{\nu}(x)\,{.}
\end{equation}
At the first step this generalized eigenvalue equation is transformed
into a standard eigenvalue equation. We use a Cholesky decomposition
\cite{PT96} of the overlap matrix
\begin{equation}
  \hat{S}=\hat{L}\hat{L}^{\rm T}
\end{equation}
into a lower triangular matrix $\hat{L}$ and its transpose and write
the eigenvalue equation as
\begin{equation}
  \hat{L}^{-1}\hat{H}\left(\hat{L}^{\rm T}\right)^{-1}\hat{L}^{\rm T}\vec{a}_{i}=\epsilon_{i}\hat{L}^{\rm T}\vec{a}_{i}\,{.}
\end{equation}
The matrix $\hat{L}^{-1}\hat{H}\left(\hat{L}^{\rm T}\right)^{-1}$ is
diagonalized using the \texttt{zheev}-routine form the LAPACK library
\cite{LAUG99} and the resulting eigenvector $\hat{L}^{\rm
  T}\vec{a}_{i}$ is transformed back to obtain the eigenvector
$\vec{a}_{i}$ of the original generalized eigenvalue problem.

The starting point for the iterative self-consistent procedure is a system of
non-interacting particles i.e. a system with $V_{\rm H}=V_{\rm x}=0$.
The resulting non-interacting eigenfunctions are then used to
construct the first approximation for the Hartee- and the exchange
potential. In the subsequent iterations the Hartree- and the
exchange potential are calculated from the eigenfunctions of the
previous step \footnote{To ensure convergence the 
  potential in the $n$-th iteration step is in fact not simply 
  calculated from the density of
  the previous step. A fraction of the self-consistent potential
  of the $(n-1)$-th step is linearly mixed to it \cite{H05}. We used the
  mixing factor $\alpha=0.2$.}. As a measure of the convergence 
we consider the maximum difference between two
Kohn-Sham eigenvalues in the $n$-th and $(n-1)$-th iteration step:
\begin{equation}
  \max_{i}\left|\varepsilon_{i}^{(n)}-\varepsilon_{i}^{(n-1)}\right|<\Delta\,{.}
\end{equation}
We found that this difference has to be extremely small compared to
the Kohn-Sham eigenvalues themselves which are of the order of several
tens of meV, namely $\Delta\approx 10^{-10}$meV. The reason for this
very small number are low energy excitations which
correspond to a charge displacement over a large distance in the
system. If the chosen $\Delta$ is too large, one encounters a density range
where the system seems to be in a delocalized state whereas in fact it becomes 
localized after the solution is converged.
Generally, a very high computational accuracy is required to distinguish
correctly between a localized and a delocalized state of the system.

\section{Results}

In this section we present the results of our calculations of the
persistent current in the one-dimensional quantum ring. For the
effective electron mass and the dielectric constant we have chosen the
GaAs values $m_{0}^{\ast}=0.0665m_{\rm e}$ and $\epsilon=12.5$. The
value of the magnetic field flux was chosen as $\Phi=0.3\Phi_{0}$.  In
fact, the particular magnitude of the flux does not matter provided the
current distinctly exceeds numerical inaccuracy.

\begin{figure}
  \includegraphics[width=0.45\textwidth]{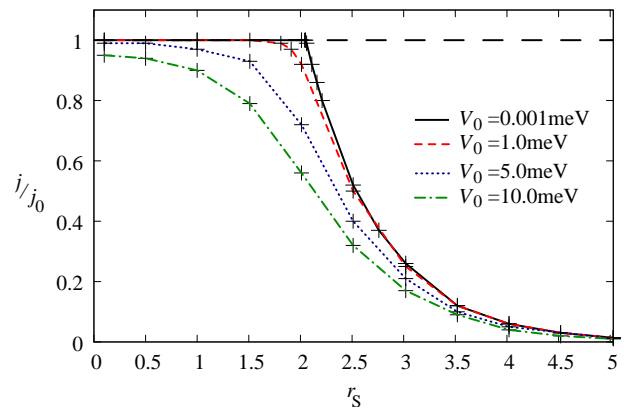}
  \caption{(Color online) The persistent current as a function of $r_{\rm S}$ for a
    Gaussian impurity potential with a half maximum width of $2.5\%$
    of the ring circumference. The current is normalized to its value
    $j_{0}$ in a non-interacting system and potential strength $V_{0}=10^{-3}$meV. 
    The long-dashed line $^j/_{j_{0}}=1$ corresponds to the interaction-free system.
    \label{fig:currden_narrow}}
\end{figure}

For the Wigner crystal pinning we apply a narrow impurity potential of
a width $\sigma=0.025 L$ much smaller than the average distance
between electrons $\frac{L}{N}=0.1 L$. The persistent current is
calculated as a function of $r_{\rm S}$, the latter being altered by
varying $m^{\ast}$, according to Eqs. (\ref{rS}), (\ref{potRenorm}).
The current is normalized to its value $j_{0}$ for non-interacting
electrons in the presence of an impurity potential with unrenormalized
strength $V_{0}=10^{-3}$meV. The results for various impurity
potential strengths are shown in Fig.  \ref{fig:currden_narrow}. The
dashed line $\frac{j}{j_{0}}=1$ reflects the current independence of
$r_{\rm S}$ for noninteracting electrons.

\begin{figure}
  \includegraphics[width=0.45\textwidth]{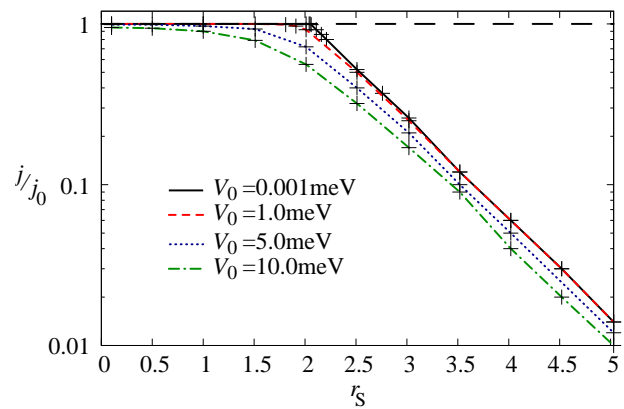}
  \caption{(Color online) Logarithmic version of the plot in Fig. \ref{fig:currden_narrow}. The exponential dependence of the persistent current on $r_{\rm S}$ is clearly seen.   
    \label{fig:currden_narrow_logscale}}
\end{figure}

As seen in Fig. \ref{fig:currden_narrow} for the smallest
$V_{0}=10^{-3}$ meV, one can clearly distinguish two different regions
of $r_{\rm S}$. Below the critical value of $r_{\rm S}^{\rm c}\approx
2.05$, the persistent current is independent of $r_{\rm S}$. Its
magnitude is the same as in the non-interacting system which means
that the interacting system is electron gas-like. In contrast, for
$r_{\rm S}>r_{\rm S}^{\rm c}$, the persistent current drops
exponentially with increasing $r_{\rm S}$ which is seen explicitely
from the linear dependence of $\log ^j/_{j_0}$ on $r_{\rm S}$ shown in
Fig. \ref{fig:currden_narrow_logscale}.  This signifies the formation
of the Wigner crystal pinned by an extremely small impurity potential.
Hence the value $r_{\rm S}^{\rm c}=2.05$ can be interpreted as a
critical $r_{\rm S}$ of the Wigner transition.

\begin{figure}
  \includegraphics[width=0.45\textwidth]{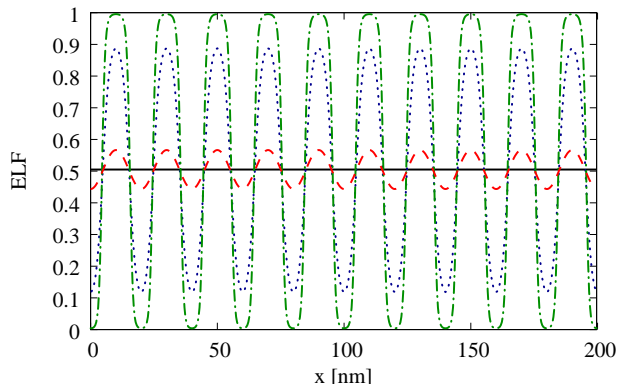}
  \caption{(Color online) Electron Localization Function in the presence of a weak
    ($V_{0}=0.001$meV) potential. Shown is the ELF for
    different values of $r_{\rm S}$. Solid line: $r_{\rm S}=0.1$,
    dashed line: $r_{\rm S}=2.06$, dotted line: $r_{\rm S}=2.5$,
    dash-dotted line: $r_{\rm S}=5.0$. An ELF value of one corresponds
    to perfect localization whereas an ELF value of one half means
    homogeneous electron gas-like delocalization.
    \label{fig:ELF_0.001}}
\end{figure}

This interpretation is supported by the ELF plot in Fig.
\ref{fig:ELF_0.001}. For $r_{\rm S} \leq 2.05$ we find an ELF value of
one half, corresponding to completely delocalized electrons. This
changes drastically when $r_{\rm S}$ exceeds $r_{\rm S}^{\rm c}$. With
increasing $r_{\rm S}$ the electrons tend to localize at discrete
lattice sites. At $r_{\rm S}\approx 5$ they arrange in an ``almost
classical'' one-dimensional lattice. The complete localization is achieved within 
a rather narrow interval of $r_{\rm S}$ as exemplified in Fig.
\ref{fig:ELF_0.001} by the ELF graphs for $r_{\rm S}=2.06$ and $r_{\rm
  S}=2.5$. This reflects the exponential decay of the persistent
current as shown in Fig. \ref{fig:currden_narrow}.

We believe, that within our numerical accuracy the solid curve in Fig.
\ref{fig:currden_narrow} corresponds to the case of the ``vanishing''
external potential. Such a potential does not disturb the Wigner
transition, but provides the pinning.  The particular potential
strength and width should be then unimportant. We tested this
calculating the current density for several values of the width of the
pinning potential (all with $V_{0}=10^{-3}$meV) and found that the
persistent current follows exactly the same $r_{\rm S}$-dependence.
However, the convergence is getting much harder for wider potentials
since the ``smoother'' potentials are less effective in pinning the
Wigner crystal. For $V_{0}$ values below $10^{-3}$meV the convergence
could not be reached. Yet using a semiclassical approach \cite{KSS95}
it can be shown analytically that the current value $j_{0}$ of a
non-interacting system is indeed recovered for $V_{\rm imp}=0$.

The critical $r_{\rm S}^{\rm c}=2.05$ we obtained in this work is of
the same order as the values for $r_{\rm S}^{\rm c}$ found in a
previous work \cite{H05} for a different model using the ground
state energy curvature \cite{Koh64} as a localization criterion. In
the presence of a disorder potential with an amplitude $\Delta
V=0.02$meV a Wigner transition has been observed in the range
$2.08\leq r_{\rm S}^{\rm c}\leq 5.04$ depending on the model for the
electron-electron interaction.

The other three curves in Fig. \ref{fig:currden_narrow} show the
current of the interacting system for $V_{0}=1.0$meV, $V_{0}=5.0$meV
and $V_{0}=10.0$meV. Although at $V_{0}=1.0$meV there is still the
range of $r_{\rm S}$ where $j=j_{\rm 0}$, the sharp kink at $r_{\rm
  S}=r_{\rm S}^{\rm c}$ vanishes. The transition smoothing is more
pronounced for $V_{0}=5.0$meV and $V_{0}=10.0$meV where no region of
$r_{\rm S}$ where the current is independent of $r_{\rm S}$ is seen.

It should be emphasized that the dependence of the normalized current
on $r_{\rm S}$ is solely due to the electron-electron interaction. The
smooth decrease of the current with increasingly strong Coulomb
interaction observed for stronger impurity potentials ($V_{0}=5.0$meV
and $V_{0}=10.0$meV) reflects a gradual localization of the many-body
state instead of a distinct phase transition. This behaviour parallels
the absence of a sharp phase transition in an external potential field
that lowers the symmetry of the high-symmetry phase \cite{Landau5}.

\begin{figure}
  \includegraphics[width=0.45\textwidth]{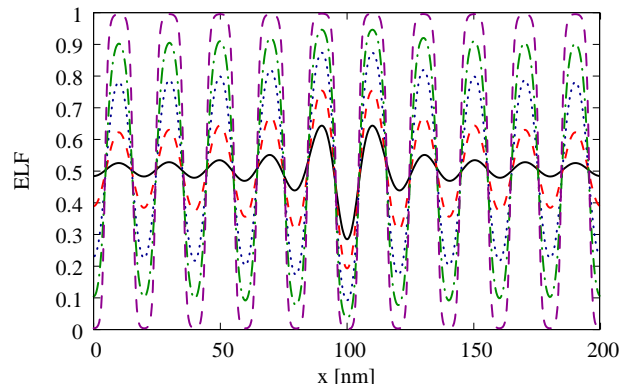}
  \caption{(Color online) Electron Localization Function in the presence of an
    intermediate ($V_{0}=5.0$meV) narrow potential. Shown is the ELF
    for different values of $r_{\rm S}$. Solid line: $r_{\rm S}=0.1$,
    dashed line: $r_{\rm S}=1.5$, dotted line: $r_{\rm S}=2.0$,
    dash-dotted line: $r_{\rm S}=2.5$, long-dashed line: $r_{\rm
      S}=5.0$. For intermediate values of $r_{\rm S}$ the electrons
    next to the impurity are more localized. This localization
    increases gradually with increasing $r_{\rm S}$. For large values
    of $r_{\rm S}$ the ELF is the same as found in the case of the
    weak potential (see Fig \ref{fig:ELF_0.001}). \label{fig:ELF_5.0}}
\end{figure}

An estimate of the Coulomb energy of two electrons at a distance
$d=\frac{L}{N}=20.0$nm 
\begin{equation}
  U=\frac{e^{2}}{4\pi\varepsilon\varepsilon_{0}}\frac{1}{d}\approx 5.75\text{meV}
\end{equation}
shows that it is indeed of the order of the pinning potential which
smoothes out the phase transition and induces a gradual localization.
For $V_{0}\ge 1$meV and at intermediate values of $r_{\rm S}$ it can
be seen directly from the ELF plots (Fig. \ref{fig:ELF_5.0}) that the
localization is more pronounced next to the pinning potential. This
indicates that a gradual localization seen in Fig.
\ref{fig:currden_narrow} is driven by the interplay between the
long-range Coulomb repulsion and the interaction with the short-range
impurity potential, both being of the same order.

\begin{figure}
  \includegraphics[width=0.45\textwidth]{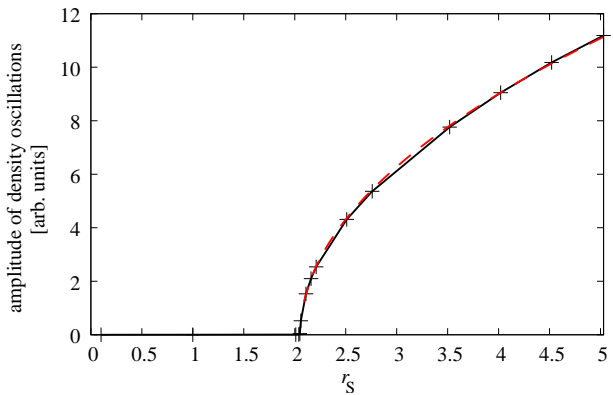}
  \caption{(Color online) Amplitude of the density oscillations as a function of
    $r_{\rm S}$ for a weak impurity potential ($V_{0}=0.001$meV). The
    solid black curve shows the calculated data, the dashed red curve is
    a square root $\left(r_{\rm S}-r_{\rm S}^{\rm c}\right)^{^1/_2}$ behaviour.
    \label{fig:order_param}}
\end{figure}

The sharp transition we found for a ``vanishing'' impurity potential
(solid line in Fig. \ref{fig:currden_narrow}) is a second order phase
transition from an electron liquid state to the Wigner crystal state.
This can be verified by plotting the $r_{\rm S}$-dependence of the
order parameter $\delta$ which shows a behaviour
$\delta\sim\left(r_{\rm S}-r_{\rm S}^{\rm c}\right)^{^1/_2}$ at
$r_{\rm S}>r_{\rm S}^{\rm c}$, i.e. in the low-symmetry phase
\cite{Landau5}.  Indeed, taking the amplitude of the density
oscillations as the order parameter $\delta$, we obtain an exact square root
dependence, as shown in Fig.  \ref{fig:order_param}. The second-order
type of the transition we observe in our calculations is quite natural
for the mean-field-type DFT-OEP approach.

From the exponential dependence of the current on $r_{\rm S}$ (Fig.
\ref{fig:currden_narrow}) we can deduce the relation between the
persistent current density and the order parameter
\begin{equation}
  j(\delta)=j_{0}\exp(-\alpha\delta^{2})
\end{equation}
where the numerical factor $\alpha=0.033L^{2}$. 

\section{Conclusions}

In this article we investigated numerically the influence of the
electron-electron interaction on the ground state of a one-dimensional
electron gas confined in a ring geometry. To break the rotational
invariance of the ring we introduce a weak ``impurity'' potential.
This potential does not affect the delocalized electron liquid phase,
but provides a pinning of the crystalline Wigner phase. We employ a
persistent current in the ring as a measure of the Wigner crystal
pinning. For a sufficiently weak impurity potential we found that for
$r_{\rm S}<r_{\rm S}^{\rm c}$ the current density of the interacting
system is exactly the same as the current density of a non-interacting
electron gas. For $r_{\rm S}>r_{\rm S}^{\rm c}$ the current of the
interacting system decays exponentially with increasing $r_{\rm S}$
while the current of a non-interacting system remains constant. This
behaviour clearly shows the formation of the Wigner crystal in a
one-dimensional system. This interpretation is confirmed by the ELF
plots which reveal the delocalized electron distribution below the
critical $r_{\rm S}^{\rm c}$ and a localized one above $r_{\rm S}^{\rm
  c}$. At $r_{\rm S}=r_{\rm S}^{\rm c}$ the system undergoes a
second-order phase transition from an electron liquid to a Wigner
crystal. This is evident from the square root dependence of the
amplitude of the density oscillations (taken as the order parameter)
on $r_{\rm S}$ above the critical value. Experimentally, this
transition should be observable as a sharp decrease of the ring's
magnetization when the electron density is lowered. However, in a real
experiment this transition will be superposed with the
interaction-independent variation of the current density due to the
variation of the particle number. The critical value $r_{\rm S}^{\rm
  c}=2.05$ we find for the Wigner transition is consistent with the
density range\cite{GRS92} in which Glazman {\it et al.}  expected the
existence of a stable one-dimensional Wigner crystal.


\begin{thebibliography}{25}
\expandafter\ifx\csname natexlab\endcsname\relax\def\natexlab#1{#1}\fi
\expandafter\ifx\csname bibnamefont\endcsname\relax
  \def\bibnamefont#1{#1}\fi
\expandafter\ifx\csname bibfnamefont\endcsname\relax
  \def\bibfnamefont#1{#1}\fi
\expandafter\ifx\csname citenamefont\endcsname\relax
  \def\citenamefont#1{#1}\fi
\expandafter\ifx\csname url\endcsname\relax
  \def\url#1{\texttt{#1}}\fi
\expandafter\ifx\csname urlprefix\endcsname\relax\def\urlprefix{URL }\fi
\providecommand{\bibinfo}[2]{#2}
\providecommand{\eprint}[2][]{\url{#2}}

\bibitem[{\citenamefont{Ihn et~al.}(2003)\citenamefont{Ihn, Fuhrer, Heinzel,
  Ensslin, Wegscheider, and Bichler}}]{IF03}
\bibinfo{author}{\bibfnamefont{T.}~\bibnamefont{Ihn}},
  \bibinfo{author}{\bibfnamefont{A.}~\bibnamefont{Fuhrer}},
  \bibinfo{author}{\bibfnamefont{T.}~\bibnamefont{Heinzel}},
  \bibinfo{author}{\bibfnamefont{K.}~\bibnamefont{Ensslin}},
  \bibinfo{author}{\bibfnamefont{W.}~\bibnamefont{Wegscheider}},
  \bibnamefont{and} \bibinfo{author}{\bibfnamefont{M.}~\bibnamefont{Bichler}},
  \bibinfo{journal}{Physica E Low-Dimensional Systems and Nanostructures}
  \textbf{\bibinfo{volume}{16}}, \bibinfo{pages}{83} (\bibinfo{year}{2003}).

\bibitem[{\citenamefont{Mailly et~al.}(1993)\citenamefont{Mailly, Chapelier,
  and Benoit}}]{MCB93}
\bibinfo{author}{\bibfnamefont{D.}~\bibnamefont{Mailly}},
  \bibinfo{author}{\bibfnamefont{C.}~\bibnamefont{Chapelier}},
  \bibnamefont{and} \bibinfo{author}{\bibfnamefont{A.}~\bibnamefont{Benoit}},
  \bibinfo{journal}{Phys. Rev. Lett.} \textbf{\bibinfo{volume}{70}},
  \bibinfo{pages}{2020} (\bibinfo{year}{1993}).

\bibitem[{\citenamefont{Lorke et~al.}(2000)\citenamefont{Lorke,
  Johannes~Luyken, Govorov, Kotthaus, Garcia, and Petroff}}]{LLG00}
\bibinfo{author}{\bibfnamefont{A.}~\bibnamefont{Lorke}},
  \bibinfo{author}{\bibfnamefont{R.}~\bibnamefont{Johannes~Luyken}},
  \bibinfo{author}{\bibfnamefont{A.~O.} \bibnamefont{Govorov}},
  \bibinfo{author}{\bibfnamefont{J.~P.} \bibnamefont{Kotthaus}},
  \bibinfo{author}{\bibfnamefont{J.~M.} \bibnamefont{Garcia}},
  \bibnamefont{and} \bibinfo{author}{\bibfnamefont{P.~M.}
  \bibnamefont{Petroff}}, \bibinfo{journal}{Phys. Rev. Lett.}
  \textbf{\bibinfo{volume}{84}}, \bibinfo{pages}{2223} (\bibinfo{year}{2000}).

\bibitem[{\citenamefont{Wigner}(1934)}]{W34}
\bibinfo{author}{\bibfnamefont{E.}~\bibnamefont{Wigner}},
  \bibinfo{journal}{Phys. Rev.} \textbf{\bibinfo{volume}{46}},
  \bibinfo{pages}{1002} (\bibinfo{year}{1934}).

\bibitem[{\citenamefont{Landau and Lifschitz}(1969)}]{Landau5}
\bibinfo{author}{\bibfnamefont{L.}~\bibnamefont{Landau}} \bibnamefont{and}
  \bibinfo{author}{\bibfnamefont{E.}~\bibnamefont{Lifschitz}},
  \emph{\bibinfo{title}{Course of theoretical physics}}, vol.
  \bibinfo{volume}{V,Statistical physics} (\bibinfo{publisher}{Pergamon Press},
  \bibinfo{address}{London}, \bibinfo{year}{1969}), \bibinfo{edition}{2nd} ed.

\bibitem[{\citenamefont{Glazman et~al.}(1992)\citenamefont{Glazman, Ruzin, and
  Shklovskii}}]{GRS92}
\bibinfo{author}{\bibfnamefont{L.~I.} \bibnamefont{Glazman}},
  \bibinfo{author}{\bibfnamefont{I.~M.} \bibnamefont{Ruzin}}, \bibnamefont{and}
  \bibinfo{author}{\bibfnamefont{B.~I.} \bibnamefont{Shklovskii}},
  \bibinfo{journal}{Phys. Rev. B} \textbf{\bibinfo{volume}{45}},
  \bibinfo{pages}{8454} (\bibinfo{year}{1992}).

\bibitem[{\citenamefont{Tanatar and Ceperley}(1989)}]{TC89}
\bibinfo{author}{\bibfnamefont{B.}~\bibnamefont{Tanatar}} \bibnamefont{and}
  \bibinfo{author}{\bibfnamefont{D.~M.} \bibnamefont{Ceperley}},
  \bibinfo{journal}{Phys. Rev. B} \textbf{\bibinfo{volume}{39}},
  \bibinfo{pages}{5005} (\bibinfo{year}{1989}).

\bibitem[{\citenamefont{Ortiz et~al.}(1999)\citenamefont{Ortiz, Harris, and
  Ballone}}]{OHB99}
\bibinfo{author}{\bibfnamefont{G.}~\bibnamefont{Ortiz}},
  \bibinfo{author}{\bibfnamefont{M.}~\bibnamefont{Harris}}, \bibnamefont{and}
  \bibinfo{author}{\bibfnamefont{P.}~\bibnamefont{Ballone}},
  \bibinfo{journal}{Phys. Rev. Lett.} \textbf{\bibinfo{volume}{82}},
  \bibinfo{pages}{5317} (\bibinfo{year}{1999}).

\bibitem[{\citenamefont{Kramer and MacKinnon}(1993)}]{KK93}
\bibinfo{author}{\bibfnamefont{B.}~\bibnamefont{Kramer}} \bibnamefont{and}
  \bibinfo{author}{\bibfnamefont{A.}~\bibnamefont{MacKinnon}},
  \bibinfo{journal}{Reports on Progress in Physics}
  \textbf{\bibinfo{volume}{56}}, \bibinfo{pages}{1469} (\bibinfo{year}{1993}),
  \urlprefix\url{http://stacks.iop.org/0034-4885/56/1469}.

\bibitem[{\citenamefont{Kohn}(1964)}]{Koh64}
\bibinfo{author}{\bibfnamefont{W.}~\bibnamefont{Kohn}}, \bibinfo{journal}{Phys.
  Rev.} \textbf{\bibinfo{volume}{133}}, \bibinfo{pages}{A171}
  (\bibinfo{year}{1964}).

\bibitem[{\citenamefont{Krive et~al.}(1995)\citenamefont{Krive, Sandstr\"om,
  Shekhter, Girvin, and Jonson}}]{KSS95}
\bibinfo{author}{\bibfnamefont{I.~V.} \bibnamefont{Krive}},
  \bibinfo{author}{\bibfnamefont{P.}~\bibnamefont{Sandstr\"om}},
  \bibinfo{author}{\bibfnamefont{R.~I.} \bibnamefont{Shekhter}},
  \bibinfo{author}{\bibfnamefont{S.~M.} \bibnamefont{Girvin}},
  \bibnamefont{and} \bibinfo{author}{\bibfnamefont{M.}~\bibnamefont{Jonson}},
  \bibinfo{journal}{Phys. Rev. B} \textbf{\bibinfo{volume}{52}},
  \bibinfo{pages}{16451} (\bibinfo{year}{1995}).

\bibitem[{\citenamefont{Sharp and Horton}(1953)}]{SH53}
\bibinfo{author}{\bibfnamefont{R.~T.} \bibnamefont{Sharp}} \bibnamefont{and}
  \bibinfo{author}{\bibfnamefont{G.~K.} \bibnamefont{Horton}},
  \bibinfo{journal}{Phys. Rev.} \textbf{\bibinfo{volume}{90}},
  \bibinfo{pages}{317} (\bibinfo{year}{1953}).

\bibitem[{\citenamefont{Talman and Shadwick}(1976)}]{TS76}
\bibinfo{author}{\bibfnamefont{J.~D.} \bibnamefont{Talman}} \bibnamefont{and}
  \bibinfo{author}{\bibfnamefont{W.~F.} \bibnamefont{Shadwick}},
  \bibinfo{journal}{Phys. Rev. A} \textbf{\bibinfo{volume}{14}},
  \bibinfo{pages}{36} (\bibinfo{year}{1976}).

\bibitem[{\citenamefont{Becke and Edgecombe}(1990)}]{BE90}
\bibinfo{author}{\bibfnamefont{A.~D.} \bibnamefont{Becke}} \bibnamefont{and}
  \bibinfo{author}{\bibfnamefont{K.~E.} \bibnamefont{Edgecombe}},
  \bibinfo{journal}{J. Chem. Phys.} \textbf{\bibinfo{volume}{92}},
  \bibinfo{pages}{5397} (\bibinfo{year}{1990}).

\bibitem[{\citenamefont{Kohn and Sham}(1965)}]{KS65}
\bibinfo{author}{\bibfnamefont{W.}~\bibnamefont{Kohn}} \bibnamefont{and}
  \bibinfo{author}{\bibfnamefont{L.~J.} \bibnamefont{Sham}},
  \bibinfo{journal}{Physical Review} \textbf{\bibinfo{volume}{140}},
  \bibinfo{pages}{A1133} (\bibinfo{year}{1965}).

\bibitem[{\citenamefont{Krieger
  et~al.}(1992{\natexlab{a}})\citenamefont{Krieger, Li, and Iafrate}}]{KLI92a}
\bibinfo{author}{\bibfnamefont{J.~B.} \bibnamefont{Krieger}},
  \bibinfo{author}{\bibfnamefont{Y.}~\bibnamefont{Li}}, \bibnamefont{and}
  \bibinfo{author}{\bibfnamefont{G.~J.} \bibnamefont{Iafrate}},
  \bibinfo{journal}{Phys. Rev. A} \textbf{\bibinfo{volume}{45}},
  \bibinfo{pages}{101} (\bibinfo{year}{1992}{\natexlab{a}}).

\bibitem[{\citenamefont{Krieger
  et~al.}(1992{\natexlab{b}})\citenamefont{Krieger, Li, and Iafrate}}]{KLI92b}
\bibinfo{author}{\bibfnamefont{J.~B.} \bibnamefont{Krieger}},
  \bibinfo{author}{\bibfnamefont{Y.}~\bibnamefont{Li}}, \bibnamefont{and}
  \bibinfo{author}{\bibfnamefont{G.~J.} \bibnamefont{Iafrate}},
  \bibinfo{journal}{Phys. Rev. A} \textbf{\bibinfo{volume}{46}},
  \bibinfo{pages}{5453} (\bibinfo{year}{1992}{\natexlab{b}}).

\bibitem[{\citenamefont{Grabo et~al.}(2000)\citenamefont{Grabo, Kreibich,
  Kurth, and Gross}}]{Gra00}
\bibinfo{author}{\bibfnamefont{T.}~\bibnamefont{Grabo}},
  \bibinfo{author}{\bibfnamefont{T.}~\bibnamefont{Kreibich}},
  \bibinfo{author}{\bibfnamefont{S.}~\bibnamefont{Kurth}}, \bibnamefont{and}
  \bibinfo{author}{\bibfnamefont{E.~K.~U.} \bibnamefont{Gross}}, in
  \emph{\bibinfo{booktitle}{Strong Coulomb correlations in electronic structure
  calculations: beyond the Local Density Approximation}}, edited by
  \bibinfo{editor}{\bibfnamefont{V.}~\bibnamefont{Anisimov}}
  (\bibinfo{publisher}{Gordon and Breach}, \bibinfo{address}{Amsterdam},
  \bibinfo{year}{2000}).

\bibitem[{\citenamefont{Vignale and Rasolt}(1988)}]{Vig88}
\bibinfo{author}{\bibfnamefont{G.}~\bibnamefont{Vignale}} \bibnamefont{and}
  \bibinfo{author}{\bibfnamefont{M.}~\bibnamefont{Rasolt}},
  \bibinfo{journal}{Phys. Rev. B} \textbf{\bibinfo{volume}{37}},
  \bibinfo{pages}{10685} (\bibinfo{year}{1988}).

\bibitem[{\citenamefont{Sharma et~al.}(2007)\citenamefont{Sharma, Pittalis,
  Kurth, Shallcross, Dewhurst, and Gross}}]{Sharma07}
\bibinfo{author}{\bibfnamefont{S.}~\bibnamefont{Sharma}},
  \bibinfo{author}{\bibfnamefont{S.}~\bibnamefont{Pittalis}},
  \bibinfo{author}{\bibfnamefont{S.}~\bibnamefont{Kurth}},
  \bibinfo{author}{\bibfnamefont{S.}~\bibnamefont{Shallcross}},
  \bibinfo{author}{\bibfnamefont{J.~K.} \bibnamefont{Dewhurst}},
  \bibnamefont{and} \bibinfo{author}{\bibfnamefont{E.~K.~U.}
  \bibnamefont{Gross}}, \bibinfo{journal}{Physical Review B (Condensed Matter
  and Materials Physics)} \textbf{\bibinfo{volume}{76}}, \bibinfo{eid}{100401}
  (pages~\bibinfo{numpages}{4}) (\bibinfo{year}{2007}),
  \urlprefix\url{http://link.aps.org/abstract/PRB/v76/e100401}.

\bibitem[{\citenamefont{Burnus et~al.}(2005)\citenamefont{Burnus, Marques, and
  Gross}}]{Burnus05}
\bibinfo{author}{\bibfnamefont{T.}~\bibnamefont{Burnus}},
  \bibinfo{author}{\bibfnamefont{M.~A.~L.} \bibnamefont{Marques}},
  \bibnamefont{and} \bibinfo{author}{\bibfnamefont{E.~K.~U.}
  \bibnamefont{Gross}}, \bibinfo{journal}{Physical Review A (Atomic, Molecular,
  and Optical Physics)} \textbf{\bibinfo{volume}{71}}, \bibinfo{eid}{010501}
  (pages~\bibinfo{numpages}{4}) (\bibinfo{year}{2005}),
  \urlprefix\url{http://link.aps.org/abstract/PRA/v71/e010501}.

\bibitem[{\citenamefont{Hofmann et~al.}(2001)\citenamefont{Hofmann, Bockstedte,
  and Pankratov}}]{HBP01}
\bibinfo{author}{\bibfnamefont{M.}~\bibnamefont{Hofmann}},
  \bibinfo{author}{\bibfnamefont{M.}~\bibnamefont{Bockstedte}},
  \bibnamefont{and}
  \bibinfo{author}{\bibfnamefont{O.}~\bibnamefont{Pankratov}},
  \bibinfo{journal}{Phys. Rev. B} \textbf{\bibinfo{volume}{64}},
  \bibinfo{pages}{245321} (\bibinfo{year}{2001}).

\bibitem[{\citenamefont{Press et~al.}(1996)\citenamefont{Press, Teukolsky,
  Vetterling, and Flannery}}]{PT96}
\bibinfo{author}{\bibfnamefont{W.~H.} \bibnamefont{Press}},
  \bibinfo{author}{\bibfnamefont{S.~A.} \bibnamefont{Teukolsky}},
  \bibinfo{author}{\bibfnamefont{W.~T.} \bibnamefont{Vetterling}},
  \bibnamefont{and} \bibinfo{author}{\bibfnamefont{B.~P.}
  \bibnamefont{Flannery}}, \emph{\bibinfo{title}{Numerical Recipes in C}}
  (\bibinfo{publisher}{Cambride University Press},
  \bibinfo{address}{Cambridge}, \bibinfo{year}{1996}).

\bibitem[{\citenamefont{Anderson et~al.}(1999)\citenamefont{Anderson, Bai,
  Bischof, Blackford, Demmel, Dongarra, Du~Croz, Greenbaum, Hammarling,
  McKenney et~al.}}]{LAUG99}
\bibinfo{author}{\bibfnamefont{E.}~\bibnamefont{Anderson}},
  \bibinfo{author}{\bibfnamefont{Z.}~\bibnamefont{Bai}},
  \bibinfo{author}{\bibfnamefont{C.}~\bibnamefont{Bischof}},
  \bibinfo{author}{\bibfnamefont{S.}~\bibnamefont{Blackford}},
  \bibinfo{author}{\bibfnamefont{J.}~\bibnamefont{Demmel}},
  \bibinfo{author}{\bibfnamefont{J.}~\bibnamefont{Dongarra}},
  \bibinfo{author}{\bibfnamefont{J.}~\bibnamefont{Du~Croz}},
  \bibinfo{author}{\bibfnamefont{A.}~\bibnamefont{Greenbaum}},
  \bibinfo{author}{\bibfnamefont{S.}~\bibnamefont{Hammarling}},
  \bibinfo{author}{\bibfnamefont{A.}~\bibnamefont{McKenney}},
  \bibnamefont{et~al.}, \emph{\bibinfo{title}{{LAPACK} Users' Guide}}
  (\bibinfo{publisher}{Society for Industrial and Applied Mathematics},
  \bibinfo{address}{Philadelphia, PA}, \bibinfo{year}{1999}).

\bibitem[{\citenamefont{Hofmann}(2005)}]{H05}
\bibinfo{author}{\bibfnamefont{M.}~\bibnamefont{Hofmann}}, Ph.D. thesis,
  \bibinfo{school}{Universit\"at Erlangen-N\"urnberg} (\bibinfo{year}{2005}).
\end{thebibliography}

\end{document}